\documentclass{article}

\usepackage[a4paper,top=2cm,bottom=2cm,left=3cm,right=3cm,marginparwidth=1.75cm]{geometry}
\usepackage[english]{babel}
\usepackage{authblk}
\usepackage{graphicx}
\usepackage{algorithm2e}
\usepackage{amsmath,amssymb}
\usepackage{dsfont}
\usepackage{multirow}
\usepackage{lineno}
\usepackage{siunitx}
\usepackage{tikz}
\usepackage{braket}
\usepackage{graphicx}
\usepackage{ragged2e}
\usepackage{xspace}

\usepackage[colorlinks=true,
            linkcolor=blue,      
            citecolor=olive,     
            filecolor=magenta,   
            urlcolor=blue]{hyperref}
\usepackage{orcidlink}

\hypersetup{
    unicode=true,
    pdftoolbar=true,
    pdfmenubar=true,
    pdffitwindow=false,
    pdfstartview={FitH},
    pdftitle={Spatially-resolved multiphoton photoemission from a lateral transition metal dichalcogenide heterostructure},
    pdfauthor={University of Oldenburg}, 
    pdfsubject={Spatially-resolved multiphoton photoemission from a lateral transition metal dichalcogenide heterostructure},
    pdfcreator={University of Oldenburg},
    pdfproducer={}, 
    pdfkeywords={},
    pdfnewwindow=true,
    colorlinks=true,
    linkcolor=blue,
    citecolor=blue,
    filecolor=magenta,
    urlcolor=blue
}

\usepackage[style=nature, sorting=none, backend=biber, doi=true,isbn=false,url=false,eprint=false]{biblatex}
\usepackage{csquotes}
\AtEveryBibitem{
  \clearlist{language}
}

\begin{document}

\newcommand{\WSe}{WSe\textsubscript{2}\xspace}
\newcommand{\MoSe}{MoSe\textsubscript{2}\xspace}

\title{Spatially-resolved multiphoton photoemission from a lateral transition metal dichalcogenide heterostructure}

\date{\today}

\author[1]{Lina Hansen}
\author[1]{Paul Martin}
\author[2]{Sai Shradha}
\author[3,4]{Julian Picker}
\author[1]{Arvid Klösgen}
\author[1]{Kerstin Harland}
\author[1]{Katrin Meier}
\author[3,5]{Andrey Turchanin}
\author[2]{Bernhard Urbaszek}
\author[1]{Jan Vogelsang\thanks{Corresponding Author: jan.vogelsang@uol.de}}
\affil[1]{Institut für Physik, Carl von Ossietzky Universität Oldenburg, 26129 Oldenburg, Germany}
\affil[2]{Institute of Condensed Matter Physics, Technische Universität Darmstadt, 64289 Darmstadt, Germany}
\affil[3]{Institute of Physical Chemistry, Friedrich Schiller University Jena, 07743 Jena, Germany}
\affil[4]{Present address: MAX IV Laboratory, Lund University, 224 84 Lund, Sweden}
\affil[5]{Abbe Centre of Photonics, 07745 Jena, Germany}

\maketitle

\begin{abstract}
Transition metal dichalcogenides (TMDs) in their monolayer form offer a premier platform for next-generation optoelectronics, particularly through the local manipulation of their robust excitonic states using nanoscale electric fields. These localized states can be dynamically controlled through spatial structuring as well as through ultrafast field modulations driven by tailored optical pulses. Characterizing the resulting rapid, nanoscale charge carrier dynamics requires a technique with exceptional spatial and temporal resolution. Here, we report on the spatio-temporally resolved investigation of ground state and excited state photoemission from a lateral heterostructure built of monolayers of \WSe and \MoSe using few-cycle light pulses with a photon energy of \SI{0.62}{eV}. We utilize photoemission electron microscopy to spatially resolve the highly nonlinear photoemission from the monolayer structure with few tens of nanometer resolution. By varying the laser pulse energy, we extract the nonlinearity of the photoemission process and thus the dynamic binding energy of the photoelectrons before and after optical excitation with high spatial and temporal resolution.
\end{abstract}

\section{Introduction}

Photoemission electron microscopy (PEEM) with down to few tens of nanometer spatial resolution has emerged as a powerful technique for probing ultrafast electron dynamics \cite{wittenbecher_unraveling_2021, zaiats_ultrafast_2026, xu_ultrafast_2024, dabrowski_ultrafast_2020, kubo_femtosecond_2005, spektor_revealing_2017, tsesses_four-dimensional_2025, li_broadband_2025}, traditionally relying on one- or two-photon processes to drive photoemission \cite{kubo_femtosecond_2005, wittenbecher_unraveling_2021, li_broadband_2025}. Recently, a growing interest has shifted toward utilizing longer excitation wavelengths to investigate field-driven charge carrier dynamics \cite{zaiats_ultrafast_2026, piglosiewicz_carrier-envelope_2014, merboldt_observation_2025, choi_observation_2025}. This development started on isolated metallic emitters such as gold or tungsten nano needles \cite{piglosiewicz_carrier-envelope_2014} and recently led to the first demonstration of Floquet engineering in monolayer graphene \cite{merboldt_observation_2025, choi_observation_2025}. Purely optical methods already benefit from highly nonlinear interactions of 2D materials with light \cite{karvonen_rapid_2017,autere_nonlinear_2018}. Using long driving wavelengths offers a distinct advantage: the ponderomotive energy of a particle in an oscillating field scales with the wavelength squared \cite{dombi_strong-field_2020} and thus long wavelengths facilitate the coherent driving of electrons via the instantaneous optical field. However, imaging photoemission experiments that use long-wavelength driving fields remain remarkably rare \cite{zaiats_ultrafast_2026}. This scarcity stems from rigorous experimental requirements, including the need for high photon fluxes, extremely short pulse durations, and high repetition rates to achieve detectable yields without detrimental Coulomb repulsion effects \cite{oloff_pump_2016, tkach_multimode_2026}. So far, the investigation of highly nonlinear photoemission processes from solids has been restricted either to spatially averaging measurements \cite{teichmann_strong-field_2015, hergert_long-lived_2017} or isolated emitters \cite{herink_field-driven_2012,piglosiewicz_carrier-envelope_2014, vogelsang_ultrafast_2015, echternkamp_strong-field_2016, schotz_nonadiabatic_2018, heimerl_quantum_2025} with only very few exceptions \cite{zaiats_ultrafast_2026}. Bridging this gap by combining long-wavelength, nonlinear excitation with high-spatial-resolution imaging holds immense promise for coherent charge carrier control in application-relevant spatially heterogeneous systems, but it is still waiting to be widely adopted.

Here we employ time-resolved photoemission electron microscopy with long-wavelength few-cycle light pulses to study local binding energy differences of charge carriers in a lateral heterostructure of a monolayer TMD. The low photon energy leads to highly nonlinear photoemission of electrons, which we characterize in detail. We use this information to extract electron binding energies with high spatial resolution and find rich spatial differences on the monolayer TMD sample. Further, by transferring this new development to a time-resolved pump-probe experiment, we show that the decay dynamics of excited states in a spatially heterogeneous sample also differ locally, which warrants a more detailed analysis. We conclude that spatially averaging investigations are not sufficient to fully describe the broad range of real-world sample systems, and ultrafast techniques with high spatial resolution are required to understand them in detail.

\section{Results and discussion}

\begin{figure*}[!htbp]
    \centering
    \includegraphics[width=0.8\textwidth]{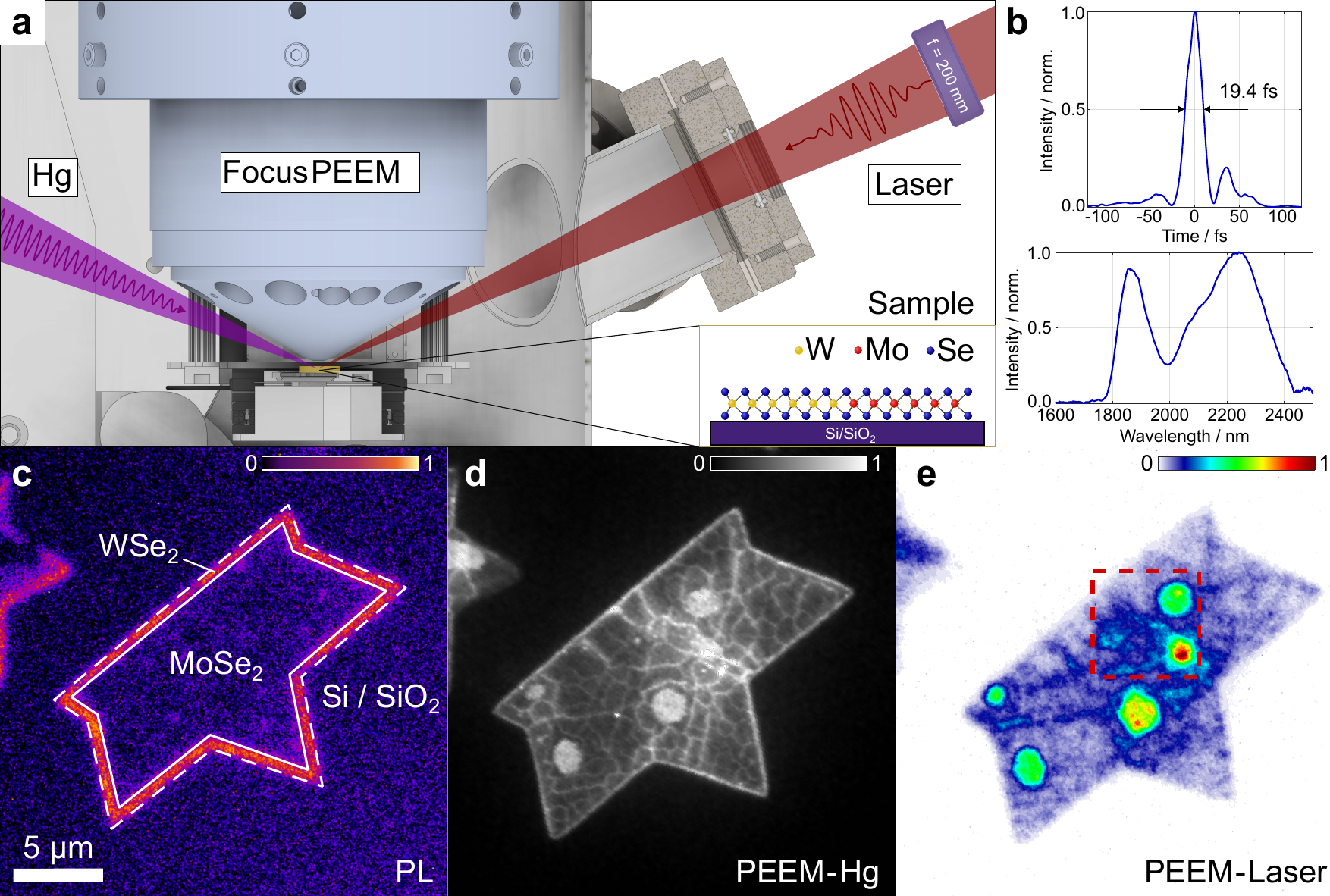}
    \caption{\justifying \textbf{a} Experimental setup of the photoemission study. The TMD monolayer sample is illuminated either by few-cycle light pulses or light from a mercury gas discharge lamp (Hg) at an angle of \SI{65}{\degree} to the surface normal. The photoelectrons are collected and analyzed in a photoemission electron microscope (PEEM). \textbf{b} Optical spectrum (bottom) of the few-cycle light pulses used for excitation and probing in the photoemission experiments. The measured temporal structure (top) reveals a pulse duration of less than \SI{20}{fs} full width at half maximum. \textbf{c} Photoluminescence (PL) measurement of the TMD monolayer heterostructure sample with a dashed outline indicating the \WSe outer region and a solid line marking the \MoSe inner region. Bright PL emission is observed from the \WSe edge region of the star-shaped structure. \textbf{d} PEEM image of the same structure recorded using a mercury gas discharge lamp as a light source for linear photoemission. Details of the sample structure are visible, including the \WSe region at the edge along with features in the central \MoSe part which are only barely visible in the photoluminescence image shown in c. \textbf{e} PEEM image recorded using photoelectrons emitted by 0.62-eV photons of a few-cycle light pulse characterized in b. A high number of six to eight photons is required for photoemission of an electron. This leads to a more dynamic image of the sample, because small differences in local field strengths and in the local work function already result in significant differences in the photoemission rate.}
    \label{fig:figure1}
\end{figure*}

We prepare a lateral heterostructure of \MoSe and \WSe monolayers using a two-step chemical vapor deposition (CVD) process. In this method, the \MoSe layer is grown first, followed by \WSe growth along the edges of the existing \MoSe monolayer domains, thereby creating a lateral heterostructure (LH) \cite{turchanin_tailored_2025, CVD2step_2021, beret2022exciton}. In this study, the resulting structure is investigated as grown on a Si substrate with a native oxide layer. Photoluminescence (PL) imaging measurements were performed at room temperature as a first characterization after the sample growth. Therefore, a home built PL microscope with a high-power white LED and 500-nm shortpass filter in the excitation beam path and 550-nm longpass filter in the detection beam path was used. The resulting PL image of the as-grown \MoSe/\WSe LH is shown in Fig.~\ref{fig:figure1}c. The brighter emission from the outer edge of the star-shaped structure is assigned to \WSe. In \WSe, the lowest-lying excitonic state is optically dark \cite{zhang2015experimental}, but thermal population of the higher-energy bright exciton at room temperature enables efficient radiative recombination and leads to higher PL emission than that from the inner \MoSe region of the LH.

We transfer the sample to the vacuum chamber of the photoemission electron microscope, operated at a pressure of \SI{1e-9}{mbar}. The low pressure both ensures stable operation of the microscope and avoids sample contamination by laser deposition of for example hydrocarbons from the residual gas. In a first step, we illuminate the sample by 4.9-eV photons from a mercury gas discharge lamp (see Fig.~\ref{fig:figure1}a). The illumination angle is \SI{65}{\degree} relative to the surface normal, limited by the geometry of the objective lens of the electron microscope. The light source provides enough flux and a sufficiently high photon energy to directly photoemit electrons from the sample and record a magnified image using the electron microscope. A medium magnification with a field of view of \SI{20}{\micro m} is chosen to get an overview of the sample and distinguish individual structures. The sample is mounted on a custom-designed two-dimensional nanopositioning system, which enables control over the region of the sample imaged by the microscope. The same part of the sample previously investigated using photoluminescence spectroscopy is identified and positioned on the optical axis of the electron microscope.

A micrograph recorded with illumination of the mercury lamp is shown in Fig.~\ref{fig:figure1}d. The silicon substrate around the star-shaped structures remains mostly dark in the image, because the number of accessible electronic states for photoemission is lower than the number of states in the star-shaped structures, leading to a weak linear photoemission signal. This makes the monolayer samples stick out clearly, with the outermost \WSe edge showing a stronger photoemission signal than the central \MoSe region. This can be well explained by the work functions $\phi$ of the two materials: The central region is expected to have a work function of approximately $\phi_\text{MoSe2} =$~\SI{4.6}{eV} compared to the edge region with a work function of approximately $\phi_\text{WSe2} =$~\SI{4.3}{eV} \cite{kim_thickness_2021}. The difference in the work function of \SI{0.3}{eV} allows in the \WSe case more electronic states to contribute to the linear photoemission process, assuming similar densities of states for the two materials. Thus, a larger fraction of the absorbed photons elevates electrons above the vacuum energy level, permitting their photoemission upon reaching the semiconductor-vacuum interface. Compared to the PL image shown in Fig.~\ref{fig:figure1}c, the \WSe edge appears narrower, which can be attributed to the better spatial resolution of PEEM compared to optical methods. Furthermore, inside the \MoSe region, a more complex structure is visible in the PEEM data. Three larger round regions with a diameter of approximately \SI{1}{\micro m} can be identified, but also darker regions with a work function closer to the work function of the silicon oxide layer on the substrate of more than \SI{5}{eV}. This indicates that the \MoSe monolayer region is not an ideal flat crystal but exhibits a more complex structure, which we investigate in more detail in the following section.

\subsection{Nonlinear ground state photoemission}

We block the illumination from the mercury lamp and direct few-cycle light pulses with a repetition rate of \SI{200}{kHz} and a central wavelength of \SI{2}{\micro m} onto the surface \cite{meier_multiscale_2026}. They are focused using a CaF\textsubscript{2} lens with a focal length of \SI{300}{mm} through a BaF\textsubscript{2} vacuum window onto the sample surface. The size of the elliptical spot on the sample surface extends over approximately \SI{70}{\micro m} by \SI{30}{\micro m} and obtains its shape due to the same illumination at a grazing incidence angle of \SI{65}{\degree} to the surface normal, just like the mercury light source before. The dispersive materials are compensated for such that a pulse duration of \SI{19.4}{fs} is reached on the sample surface (see Fig.~\ref{fig:figure1}b). Photoelectrons are emitted from the sample surface despite the low mean photon energy of only \SI{0.62}{eV}. The ultrashort pulse duration in combination with a moderate pulse energy of \SI{20}{nJ} and a rather tight focus of the light onto the sample surface provides a sufficiently high intensity to drive nonlinear photoemission processes with measurable probability. Classically speaking, a number of photons $n$ is absorbed quasi-simultaneously and provides the required energy for an electron to overcome the vacuum barrier. The resulting spatial emission pattern of photoelectrons is shown in Fig.~\ref{fig:figure1}e with the star-shaped structure again visible. The round regions are even more prominent compared to the image shown in Fig.~\ref{fig:figure1}d, while the edge region consisting of \WSe is barely visible. 

\begin{figure*}[!htbp]
    \centering
    \includegraphics[width=0.9\textwidth]{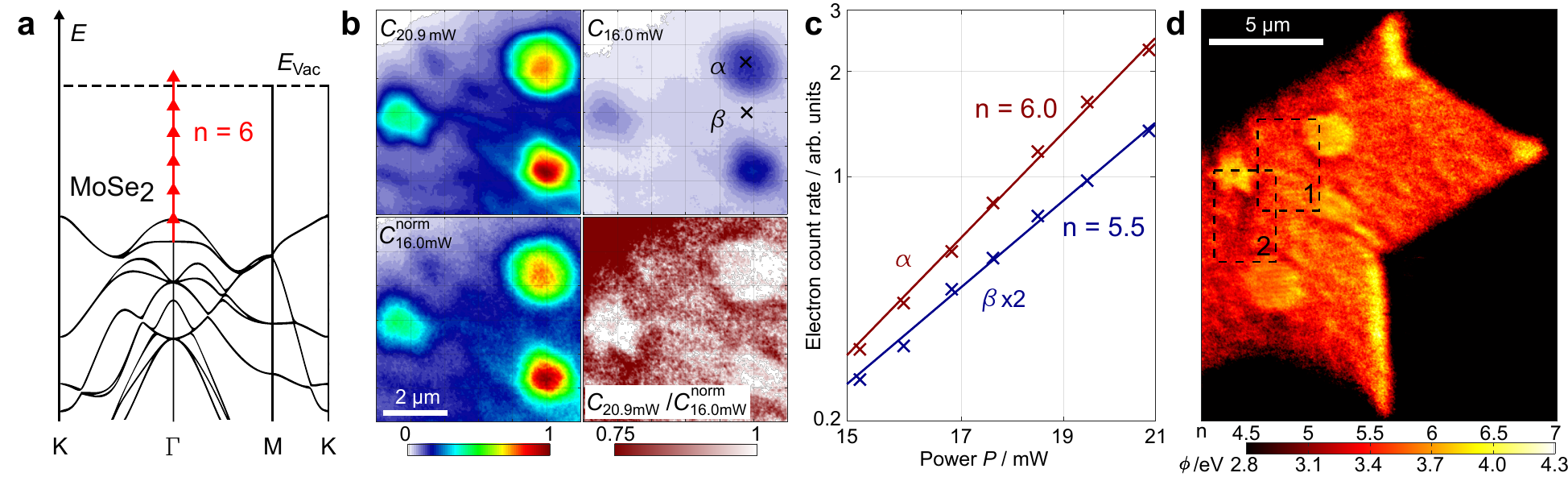}
    \caption{\justifying \textbf{a} Schematic of the multiphoton photoemission process. A high number of photons $n$ is required to supply enough energy to an electron to overcome the potential barrier $\phi$, followed by photoemission upon reaching the surface. \textbf{b} Photoemission signal from the marked region in Fig.~\ref{fig:figure1}e for two different excitation powers $P_7 = 20.9$ mW and $P_2 = 16.0$ mW. In the bottom right, the images recorded with the two laser powers are compared. \textbf{c} Local determination of nonlinearity. By plotting the count rate for the marked points $\alpha$ and $\beta$ from b against the average laser power in a double-logarithmic plot, followed by a polynomial fit, the nonlinearity of the emission process for each point can be determined from the slope of the curve. \textbf{d} Calculating the nonlinearity at each pixel generates a map of nonlinearities, which shows clear spatial differences. The difference between the substrate (values of less than 5) and the lateral heterostructure (values between 5 and 6.5) is clearly evident. More subtle differences are visible inside the monolayer material and two regions of interest are marked that will be investigated in more detail. The nonlinearity of photoemission can be used to calculate the local work function $\phi$ of the material, as described in the main text. The color bar has been supplemented accordingly.}
    \label{fig:figure2}
\end{figure*}

Considering the theoretical photon energy and the work function $\phi$ of the materials under investigation, a number $n$ of 7 to 8 photons is required, making the photoemission process highly nonlinear (see the band diagram in Fig.~\ref{fig:figure2}a). This in particular means that an intensity $I$, locally enhanced by a factor $k$ would lead to an increase in the emission rate by a factor $k^n$. Thus, a local intensity enhancement, for example due to the geometry, of just \SI{20}{\percent} can increase the emission signal of an 8-photon photoemission process by a factor of 4. This is reflected in the emission image in Fig.~\ref{fig:figure1}e, which shows less of the overall structure visible in the mercury lamp image, but more localized emission from regions with locally enhanced fields. Thus, the excitation with a low photon energy (a long wavelength) enhances the spatial selectivity in the photoemission process, restricting photoemission to ultrasmall spots that can be significantly smaller than the wavelength. A broad range of colors is used in the color map to partially compensate for the broad dynamic range of the image.

Only faintly visible in certain regions of Fig.~\ref{fig:figure1}e is the material contrast between the inner \MoSe region and the \WSe edge. In the top right corner and partially also in the bottom left corner, a marginally enhanced electron emission count can be observed at the edge region compared to the adjacent more central region. The observation is hence similar to the Hg image, although the explanation is slightly more complex. Taking into account the literature values of the work function of approximately \SI{4.6}{eV} and \SI{4.3}{eV} for the central and edge regions, respectively, and a laser photon energy of \SI{0.62}{eV}, eight and seven photons would be needed to emit an electron into the vacuum, respectively. This should result in a sizable difference in electron count rate, which is not observed. Local differences in the environment of the sample will likely alter the work function of the material, potentially leading to a situation where an equal number of photons is needed for both materials for photoemission. This is why we will investigate the local work functions in the following section using the photoemission nonlinearity.

To look closer into the highly nonlinear photoemission process for different spatial regions, we change the laser pulse energy illuminating the sample. This directly affects the photoemission rate as the pulse energy is proportional to the local intensity on the sample. We vary the average laser power $P$ from 15.2 to \SI{20.9}{mW}, corresponding to pulse energies of 76 to \SI{104}{nJ}. The measured photoemission signals at the highest power $C_{20.9\text{mW}}$ and the second lowest $C_{16.0mW}$ are shown in Fig.~\ref{fig:figure2}b. In the top row, both images are displayed with the correct relative count rate. As expected, the lower laser power leads to a significant reduction in electron emission rate by a factor of approximately five. By normalizing the top right image $C_{16.0\text{mW}}$ to its maximum value just as the top left image, we can compare the individual features visible in the images taken at lower and higher laser power on the left. Differences are barely visible except for a slight increase in relative signal amplitude in the features around the three main emission spots for the lower laser power compared to the image taken at maximum power.

To better compare the relative emission rates, the values of both normalized images are divided by each other, and the resulting image with relative emission rates is presented in the bottom right part of Fig.~\ref{fig:figure2}b. The three prominent white regions have a ratio of approximately one. With respect to the bottom right region, this is not surprising as both images had been normalized to this area, and thus a ratio of 1 is by definition the outcome. Both of the other two regions appear also white in the bottom right part of the figure, although their signal amplitude is lower. This shows that the relative count rate on all three spots is independent of the laser power and thus the same nonlinearity leads to photoemission. The difference in absolute count rate can thus only be explained by a difference in the available states or differences in the local field enhancement in the respective areas. In contrast, the regions around the three main emitters show a more dynamic pattern. First of all, the relative count rate in this area is higher at lower laser power than at higher laser power, visible by the more reddish color. Thus, the emission rate does not grow as fast in this area as in the three hot spot regions, indicating a lower nonlinearity of the photoemission process. Apart from this behavior on a multi-micrometer scale, there are additional features recognizable on the sub-micrometer scale.

To investigate the small local features more closely and to quantify the differences in the photoemission process, we first look at two positions on the sample exemplarily. We take a closer look at the marked positions $\alpha$ and $\beta$ and plot the photoemission rates $C$ for seven different laser powers in Fig.~\ref{fig:figure2}c. As before, we observe an increase in the electron count rate with laser power. In the chosen double-logarithmic representation, the increase follows a linear trend for both positions with a small difference in slope. The linear slope is expected as we observe a nonlinear photoemission process of order $n$. With $p$ being the probability of a single photon absorption event, the emission rate $C$ for a process of order $n$ scales with $C\propto p^n$. Furthermore, the probability of an individual event $p$ is proportional to the number of photons available and thus is proportional to the laser power $P$: $c\propto P^n$.

Power-law least-squares fits to both datasets reveal a nonlinearity $n$ of 6.0 for $\alpha$ and a slightly lower nonlinearity of 5.5 for $\beta$. This confirms the previous observation that the nonlinearity of the emission from the hot spots is higher than the nonlinearity of photoemission from the surrounding region. With this analysis, we can quantify the difference and calculate the effective local work function $\phi$ from the measured nonlinearities of photoemission. Taking into account the average photon energy of \SI{0.62}{eV}, the measured nonlinearities correspond to a work function of \SI{3.72}{eV} and \SI{3.41}{eV} for the two areas in the \MoSe region, which is slightly lower than the literature value for \MoSe of \SI{4.3}{eV}. A redshift of similar energy difference was observed in PL measurements of exciton peaks of \MoSe and \WSe in lateral heterostructures of the same type, which can be attributed to tensile strain resulting from the growth process on a Si substrate \cite{Hossain_CVD-grown_2026}. Additionally, the extraction field of \SI{12}{kV} of the microscope's objective lens alters the binding potential via the Schottky effect, which leads to a small reduction of the measured work function. Another major contribution, apart from contaminations of the sample, can be the locally different coupling of the monolayer to the silicon substrate, which leads to band renormalization, and thus also influences the measured work function.  

The analysis method introduced in the previous paragraph is now applied to each individual pixel in the image of the monolayer sample. A series of images was recorded over several hours to collect the required amount of data for such a fine-grained analysis. The resulting nonlinearity for each detector pixel is drawn as an image in Fig.~\ref{fig:figure2}d. We find a distribution of nonlinearities ranging from 5 to 7 for the monolayer region and lower values for the surrounding bare substrate. Again, the hot spot regions stick out with a particularly high nonlinearity of up to 6.5 as exemplarily studied for point $\alpha$ before. Similarly, part of the edge region shows a higher nonlinearity as well, in particular on the right hand side of the star shaped structure. This cannot be explained with the work function difference between the edge and the central region. The image taken with the mercury lamp revealed a clearly higher emission rate from the edge region, indicating the already discussed lower work function, which is also expected from literature. The most likely explanation is a geometric effect due to the laser illumination at an angle of \SI{65}{\degree} to the surface normal from the right hand side of the image. The field strength can be enhanced locally at the illuminated edge region, leading to a more efficient ultrafast excitation of charge carriers to higher states, forming a thermal hot electron distribution via carrier carrier scattering. Emission from these transiently occupied states can then be driven more efficiently. The temporal dynamics will be discussed more in the last part of this work.

\begin{figure*}[!htbp]
    \centering
    \includegraphics[width=0.9\textwidth]{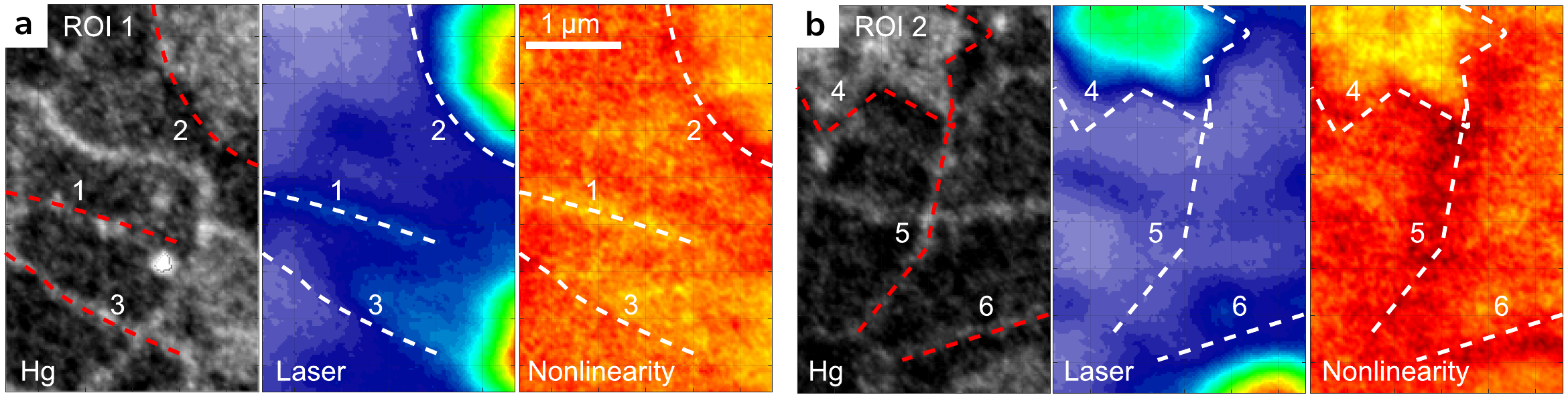}
    \caption{\justifying \textbf{a} and \textbf{b} Detailed analysis of the two regions 1 and 2 marked in Fig.~\ref{fig:figure2}d, respectively. A comparison between the photoemission signal in PEEM through excitation by a mercury lamp (Hg, left), the signal caused by laser excitation (laser, middle) and the calculated nonlinearity values (nonlinearity, right) of regions 1 and 2 is shown. The dashed lines 1-6 highlight areas with particularly prominent patterns. Their differences can be explained by different mechanisms involved in the emission process, which are discussed in the main text.}
    \label{fig:figure3}
\end{figure*}

In the next step, we discuss the rich structure visible in Fig.~\ref{fig:figure2}d in more detail. For this, we zoom into the two marked regions with particularly diverse behavior, utilizing the spatial resolution of our method that is better than \SI{100}{nm} (see SI). To this end, we directly compare the three information channels we have available: Emission driven by the mercury lamp, the nonlinear photoemission signal after laser illumination, and the nonlinearity of the emission derived as discussed before. Fig.~\ref{fig:figure3} shows the results for both regions of interest. We find distinctly different behavior in different areas of the sample, marked by dashed lines in Fig.~\ref{fig:figure3}.

The area surrounding dashed line 1 has high values for all three representations, which means a high photoemission signal for Hg and laser excitation and a high nonlinearity value  of $n \approx$ 6. The high nonlinearity observed here indicates a comparably high binding energy of the photoelectrons before emission. Yet, the simultaneously high signal in the mercury lamp image allows us to conclude that the density of states at this energy is increased compared to that of the surroundings, permitting a strong emission signal despite the reduced energy range accessible by the photons of the mercury lamp due to the higher binding energy. The laser signal shown in the middle confirms this, as also the photoemission rate is slightly increased.

In direct contrast, the area underneath the dotted line 2 shows lower values for all three representations. Thus, the opposite conclusion can be drawn: There are states available at low binding energy, hence the low nonlinearity ($n\approx$ 5.2). But their density is comparably low, leading to a reduced laser emission signal and also a low signal using mercury lamp illumination.

Although high values can be seen in the Hg image near the third line, no distinct structure is visible in the laser signal, and a step edge is present in the nonlinearity map. The step from higher nonlinearity to lower nonlinearity indicates a change in the electronic configuration of the sample, potentially due to a structural change in the material itself or a change in interaction with the substrate. The line visible in the Hg image  accordingly marks the interface and could show an enhanced emission for geometric reasons as the Hg light source illuminates the sample from the top right. In contrast, the laser illuminates the sample from the side and thus does scatter differently with the step edge, leading to a strongly reduced visibility of the feature in the laser emission image in the middle.  

In the second region of interest, shown in Fig.~\ref{fig:figure3}b, the area enclosed by the dashed line 4 behaves similarly to area 1 in the first region of interest discussed before. Additionally, it shows an interesting sub-structure, which is not visible in the laser emission image in the middle, but in the Hg image and, in particular, in the nonlinearity image. This demonstrates quite nicely how the change in laser power and the deduced nonlinearity provide additional information about the sample, such as local work function differences.  

As a last special feature, the areas marked by lines 5 and 6 show high values in the Hg representation, whilst they have no significant contribution to the laser signal, but they are clearly visible in the nonlinearity map with low values ($n\approx$ 5.2). The low nonlinearity agrees well with the Hg lamp signal as several states with low binding energy are available for photoemission. The missing contrast in multiphoton photoemission with the laser pulses indicates that the accessible density of states for photoemission does not differ from the surrounding area. This means that while electrons are bound more strongly, they are available with a higher density, which compensates for the lower emission probability.

Overall, this analysis clearly points out that evaluating the nonlinearity map and comparing it with the photoemission signals from Hg and laser excitation allows us to draw conclusions about the emission process at spatially well-defined points that were previously inaccessible. 

\subsection{Nonlinear excited state photoemission}

\begin{figure*}[!htbp]
    \centering
    \includegraphics[width=0.6\textwidth]{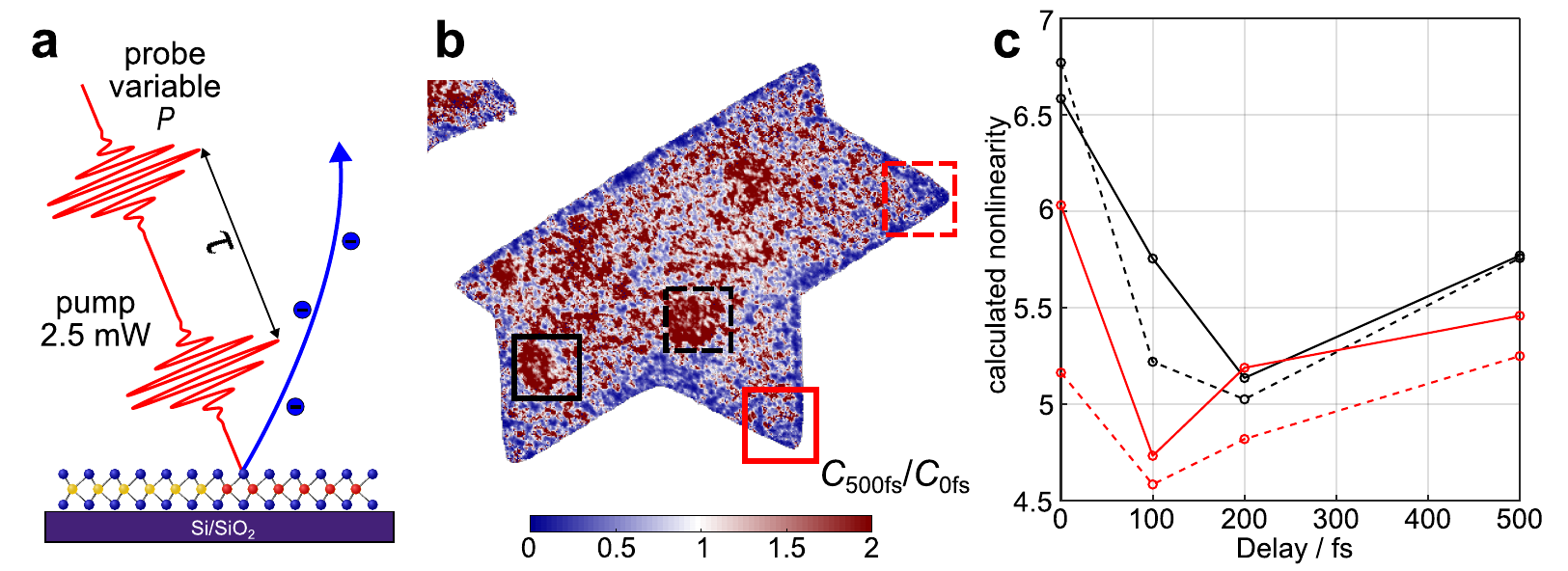}
    \caption{\justifying \textbf{a} Schematic of the time-resolved experiment. A first pulse with a constant energy of \SI{12.5}{nJ} excites the TMD monolayer sample and a second pulse probes its state after a variable delay $\tau$ with variable pulse energy. \textbf{b} Quotient image of the static and the dynamic emission signal. Both images have been normalized before dividing the local values of the dynamic image by the local values of the static image. The \WSe region clearly appears blue, indicating that the signal from this region has decayed significantly faster than the remaining signal in the central \MoSe region. Additionally, local differences are visible. \textbf{c} Time-dependent nonlinearities, determined as before, at four different time delays for the marked rectangular regions in b. A clear reduction in nonlinearity can be observed for small time delays followed by an increase. The minimum of the measured nonlinearity is reached at $\tau\approx 100\,\mbox{fs}$ for the red regions at the star edges compared to about \SI{200}{fs} in the square regions located more towards the center. } 
    \label{fig:figure4}
\end{figure*}

Following the previous static investigations of local nonlinearities, we now investigate how the nonlinearity of photoemission changes after optical excitation and a variable waiting time $\tau$. For this, we first excite the sample with a 2.5-mW excitation pulse (see also Fig.~\ref{fig:figure4}a, corresponding to a pulse energy of \SI{12.5}{nJ}), which leads only to a weak photoemission signal. After a waiting time $\tau$, which is set via a Mach-Zehnder interferometer in a range of \SI{-100}{fs} to \SI{500}{fs}, a second pulse arrives with variable pulse energy. We vary its pulse energy as before to determine the nonlinearity of the photoemission process. Since this set of measurement data was recorded under slightly different focusing conditions than before, the required power values are lower than before.  

As previously in the static case, we will first examine the difference in the relative count rates to directly observe spatial changes. Therefore, we normalize the dynamic emission image for a time delay of \SI{500}{fs} and divide it by the static reference image  previously shown in Fig.~\ref{fig:figure1}e. In addition, a mask was created from the Hg image shown in Fig.~\ref{fig:figure1}d to cut out the substrate around the structures, which forms a noisy background in the quotient image due to the low signal level (see SI for more details). The result is shown in Fig.~\ref{fig:figure4}b, in which blue areas indicate a lower relative signal strength in the dynamic image, while red areas indicate a higher relative signal. 

The blue areas are mainly found at the edges of the sample, and thus precisely match the \WSe regions of it. The red areas are located more towards the center and in particular in the circular regions already previously identified in the static mercury lamp image shown in Fig.~\ref{fig:figure1}d. Before we interpret this observation, we have a closer look at the local nonlinearities for different time delays. We exemplarily select four regions on the sample (squares in Fig.~\ref{fig:figure4}b) and draw the experimentally determined nonlinearities in the areas for different time delays (Fig.~\ref{fig:figure4}c). We observe distinctly different evolutions of the nonlinearities depending on the spot on the sample and find both differences in decay times, recovery times, and strengths of these observations. Comparing red and black squares, i.e. edge regions versus central regions, we find a faster and deeper decay of the observed nonlinearity, indicating a faster build-up of a higher electronic excitation at the edges compared to the center. The recovery seems to be faster as well towards the edges, i.e. a faster decay of the excited charges back to their ground state. For example, the red dashed area is fully recovered after \SI{500}{fs}, while the signal from the black dashed area still remains approximately \SI{15}{\percent} below the static nonlinearity level.

We find it surprising that for some positions, in particular towards the central region, a minimum of the nonlinearity and thus a maximum of the excited state population is reached after a rather long delay time of \SI{200}{fs}. The excitation pulse with a duration of only \SI{20}{fs} cannot be responsible and thus a transfer of charge carriers to a state that can be easily probed by the probe pulse must occur during this initial time period. This could either happen as a spatial charge transfer, for example from the substrate to the surface of the TMD monolayer, or as a transfer in momentum space towards states closer to the $\Gamma$ point, which we probe with our photon energy.

Coming back to the quotient image in Fig.~\ref{fig:figure4}b, the blue edge can be seen as a faster decay of excited charge carriers to their ground state compared to the \MoSe region. The longer-lived excited state charge carriers in the \MoSe region enhance the signal relative to the edge by a lower nonlinearity and thus a higher photoemission probability. Thus, the observed nonlinearity of photoemission becomes a sensitive probe not only for the local work function differences on a real, imperfect sample, but also allows us to track the dynamics of excited charge carriers.

\section{Conclusion}

In summary, we have imaged the lateral heterostructure of a TMD monolayer for the first time using photoemission electron microscopy. A rich structure is visible upon illumination by a mercury lamp, clearly showing the lateral interface between the two materials as well as additional structures related to the fabrication of the sample and potential contamination present on the sample surface. For the first time, we study the highly nonlinear photoemission from a TMD sample using 2-µm few-cycle light pulses, which require a total number of 6 to 8 photons to photoemit an electron. Using the high spatial resolution of PEEM, we find distinct local differences in the nonlinearity of photoelectron emission and thus in the binding energy of the electrons prior to light absorption. We utilize this new way of probing charge carriers in a time-resolved experiment to investigate excitation and decay dynamics locally in the 2D material with high spatial and temporal resolution. It turns out that the nonlinearity is a quite sensitive measure of the binding energy of an electron, in particular for the low photon energies of \SI{0.62}{eV} used here. We are convinced that a wavelength of \SI{2}{\micro m} is quite ideal for this purpose as it simultaneously provides the discussed energy resolution and still allows for short pulse durations of less than \SI{20}{fs}. This combination provides the opportunity to probe dynamics with high spatial, temporal and energy resolution as well as field-driven control of dynamics using the long-wavelength field in a next step.

\section*{Acknowledgments}
We would like to thank Germann Hergert for helpful discussions.

\section*{Data availability statement}
The data that support the findings of this study are available upon reasonable request from the authors.

\section*{Funding}
J.P. and A.T. acknowledge the financial support of this work via the Deutsche Forschungsgemeinschaft (DFG, German Research Foundation) SPP 2244 ‘2DMP’ (Project TU149/21-1, 535253440), and DFG individual grant TU149/16-1 (464283495). S.Sh. and B.U. acknowledge financial support by the
DFG via SPP 2244. J.V. acknowledges support by the zu\-kunft.nie\-der\-sach\-sen program of the Niedersächsisches Ministerium für Wissenschaft und Kultur (DyNano and Stay Inspired) and the DFG (462448709, Emmy Noether program).


\printbibliography

\end{document}